# Quantum Networks

by

Bassam Aoun

&

Mohamad Tarifi

## ABSTRACT

We first give a brief overview over quantum computing, quantum key distribution (QKD), a practical architecture that integrates (QKD) in current internet security architectures, and aspects of network security. We introduce the concept of quantum contracts inspired from game theory. Finally, we introduce the basic architecture of the quantum internet and present some protocols.

# ACKNOWLEDGEMENTS

*We would like to thank Prof. Alex Lopez-Ortiz for his accessibility that made this possible.*

*We would like to thank Peter Szulczewski, Piragash Velummylun, and Khosrow Ebrahimpour for there company and encouragement.*

*We also thank University of Waterloo for being a good host.*

# CONTENTS

# INTRODUCTION

Quantum mechanics is our current best description of the world as we know it. Experiments show quantum predictions where accurate up to 10-decimal places. In quantum cryptography much work has been devoted to the study of Quantum Key Distribution (QKD). The purpose of QKD is to securely distribute secret keys between users in a network. The result of this investigation was several quantum protocols that have been later implemented and tested [5, 6]. The first of such protocols was the BB84 due to Bennett and Brassard [7], a later version of this protocol was proved to be unconditionally secure. We will discuss this and other QKD protocols. Other intriguing application of quantum mechanics is in the solution of other network security and distributed computing problems such as the Byzantine Agreement[i] and Fingerprinting[ii]. We will attempt to give a brief account for the proposed quantum solutions for those and other problems in the third section. Next we relate some of the results from quantum games to network security and cryptography; in particular we develop the notion of Quantum Contracts (QCNTs), which where hinted upon by Benjamin and Hayden [iii] in their study of multiplayer quantum games. We will attempt to give a definition for QCNTs and explore its use in the context of a network.

# OVERVIEW OF QUANTUM INFORMATION

## THE BEAM SPLITTER EXPERIMENT

As a start, we will illustrate some counterintuitive concepts of quantum mechanics through an experiment. Those concepts seem counterintuitive because everyday phenomenons are governed by classical physics, not quantum mechanics -- which takes over at the atomic level.

In figure 1, a light source emits a photon along a path towards a half-silvered mirror. This mirror splits the light, reflecting half vertically towards detector A and transmitting half toward detector B. Our intuition would say that the photon leaves the mirror either towards A or B with equal probability since it cannot be split. The fact that a photon cannot split have been verified through detecting a signal at only one detector. This means that photons will be detected 50% of the time at each of the two detectors. So far, the quantum physical prediction agrees with the classical one.

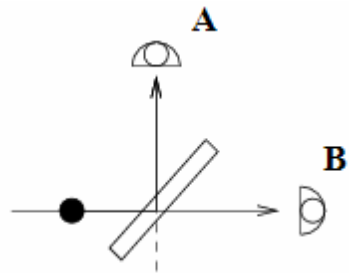

**Figure 1: Experiment 1 using one beam splitter**

This peace of information is misleading since it might lead us to think that the photon leaves either towards A or towards B. However, quantum mechanics predicts, through the effect known as *single-particle interference*, that the photon actually travels along both paths simultaneously, collapsing down to one path only upon measurement. The following experiment illustrates the last effect.

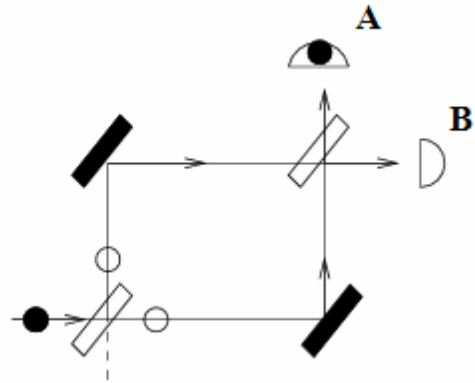

**Figure 2: Experiment 2 using two beam splitters**

In this experiment, we introduce a fully silvered mirror instead of each detector of the first experiment such that the two paths will encounter a half-silvered mirror before reaching detectors A and B. Once a photon will reach the last half-silvered mirror, along either one of the two paths, one might think that the photon will reach the detectors A or B with a probability of 50% each. However in this experiment, the detector A or the detector B will register a hit 100% of the time whereas the other one will never be triggered.

In this experiment, our classical intuition based on the conditional probability doesn't predict such outcome. We cannot explain this conclusion based on a comparison with the first experiment. This phenomenon is known as single-particle *interference.* Actually, quantum physics states that the photon traveled both paths simultaneously; creating interference at the point of intersection that canceled the possibility of the photon to reach the other detector. Consequently, if we cancel out the effect of quantum interference by placing an absorbing screen on one of the paths, both detectors will registers 50% hits similar to the first experiment. Those potential paths taken by the photon represent the superposition of the possible photon states.

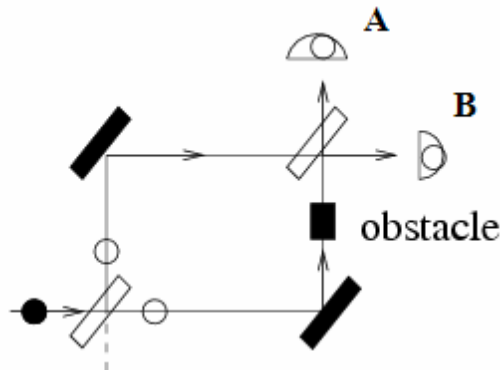

**Figure 3: Placing an obstacle on one of the paths**

Those special characteristics as the superposition of different states and interference give the quantum computer the potential to be incredible powerful computational devices. Therefore, quantum computers are not seen as continuity of classical computers but as an entirely new branch of thoughts.

QUANTUM MECHANICS

In this section we introduce the four postulates of quantum mechanics as they are relevant to our investigation in quantum information processing. Quantum postulates are very important in a sense that they provide the connections between the physical, real, world and the quantum mechanics mathematics used to model these systems.

**Postulate 1**: *Any isolated physical space is associated with a complex vector space with inner product called the State Space of the system.* It states that a system is completely described by a state vector, a unit vector, pertaining to the state space which describes all the possible states the system can be in.

**Postulate 2:** *Evolution of an isolated system can be as:*

$$|v(t_2)\rangle = U(t_1, t_2)|v(t_1)\rangle$$

*where t1, t2 are moments in time and U(t1, t2) is a unitary operator*. We should note that the process is in a sense Markovian (history doesn't matter) and reversible, since

$$U^\dagger U |v\rangle = |v\rangle$$

**Postulate 3**: *The measurement of a closed system is described by a collection of operators Mm which act on the state space such that*

- $\rho(m) = \langle \psi | M_m^\dagger M_m | \psi \rangle$ *describes the probability the measurement outcome* **m** *occurred,*

- $|\psi'\rangle = \dfrac{M_m|\psi\rangle}{\sqrt{\langle\psi|M_m^\dagger M_m|\psi\rangle}}$ is the state of the system after measurement outcome **m** occurred,

- $\sum_m M_m^\dagger M_m = I \Leftrightarrow \sum_m p(m) = 1$ *(Completeness relation)*.

Note that measurement is an external observation of a system and so disturbs its unitary evolution.

**Postulate 4:** *The state space of a composite system is the tensor product of the state spaces of its components*

$$\left.\begin{array}{l} System\ A:\ |x\rangle \\ System\ B:\ |\psi\rangle \end{array}\right\} System\ AB:\ |x\rangle \otimes |\psi\rangle.$$

In common usage, $|x\rangle$ represents the physical system whereas $|\psi\rangle$ represents the ancilla system (corresponds to measurement outcomes). This leads to:

U*nitary Dynamics + Projective Measurements = General Measurements*

QUBITS

The fundamental resource and basic unit of quantum information is the quantum bit (qubit). From a physical point of view, a qubit is represented by an ideal two-state quantum system. Examples of such systems include photons (vertical and horizontal polarization), electrons and other spin-1/2 systems (spin up and down), and systems defined by two energy levels of atoms or ions. From the beginning the two-state system played a central role in studies of quantum mechanics. It is the most simple quantum system, and in principle all other quantum systems can be modeled in the state space of collections of qubits.

A qubit is represented as unit vector in a two dimensional complex vector space for which a particular orthonormal basis, denoted by $\{|0\rangle, |1\rangle\}$, has been fixed. The notation for these states was introduced by Dirac. It is called the "ket" notation, and its variations are widely used in quantum physics. It is important to notice that the basis vector $|0\rangle$ is not the zero vector of the vector space.

For the purposes of quantum computing, the basis states $|0\rangle$ and $|1\rangle$ encode the classical bit values 0 and 1 respectively. Unlike classical bits however, qubits can be in a superposition of $|0\rangle$ and $|1\rangle$ such as $\alpha|0\rangle + \beta|1\rangle$ where α and β are complex numbers such that $|\alpha|^2 + |\beta|^2 = 1$. If such a superposition is measured with respect to the basis $\{|0\rangle, |1\rangle\}$, the probability that the measured value is $|0\rangle$ is $|\alpha|^2$ and the probability that the measured value is $|1\rangle$ is $|\beta|^2$.

Key properties of quantum bits:

1. A qubit can be in a superposition state of 0 and 1.

2. Measurement of a qubit in a superposition state will yield probabilistic results.

3. Measurement of a qubit changes the state to the one measured.

TENSOR PRODUCTS

Much computational power of quantum systems comes from the fact that as the number of qubits increases linearly, the amount of information stored increases exponentially. For example, a single-qubit state $|\psi\rangle \in \mathbb{C}^2$ is represented by a pair of complex numbers: $|\psi\rangle = \alpha|0\rangle + \beta|1\rangle$. The composite state of two qubits is an element of $\mathbb{C}^4$:

$$\alpha_{00}|00\rangle + \alpha_{01}|01\rangle + \alpha_{10}|10\rangle + \alpha_{11}|11\rangle.$$

The composite state of three qubits is in $\mathbb{C}^8$, and so on.

More generally, if $H_1$ and $H_2$ are Hilbert spaces, then $H_1 \otimes H_2$ is also a Hilbert space. If $H_1$ and $H_2$ are finite dimensional with bases $\{u_1, u_2, ... u_n\}$ and $\{v_1, v_2, ... v_n\}$ respectively, then $H_1 \otimes H_2$ has dimension *nm* with basis $\{u_i \otimes v_j \mid 1 \leq i \leq n, 1 \leq j \leq m\}$.

For matrices $A$, $B$, $C$, $D$, $U$ and scalars *a, b, c, d* the following hold:

$$\begin{pmatrix} A & B \\ C & D \end{pmatrix} \otimes U = \begin{pmatrix} A \otimes U & B \otimes U \\ C \otimes U & D \otimes U \end{pmatrix} \quad \text{and} \quad \begin{pmatrix} a & b \\ c & d \end{pmatrix} \otimes U = \begin{pmatrix} a \otimes U & b \otimes U \\ c \otimes U & d \otimes U \end{pmatrix}$$

The tensor product of several matrices is unitary if and only if each one of the matrices is unitary up to a constant. Let $U = A_1 \otimes ... \otimes A_n$. Then $U$ is unitary if $A_i^\dagger A_i = k_i I$ and $\prod_i k_i = 1$.

$$U^\dagger U = (A_1^\dagger \otimes ... \otimes A_n^\dagger)(A_1 \otimes ... \otimes A_n) = A_1^\dagger A_1 \otimes ... \otimes A_n^\dagger A_n = k_1 I \otimes ... \otimes k_n I = I$$

We can define an inner product on $U \otimes V$ by

$$(|v_1\rangle \otimes |u_1\rangle, |v_2\rangle \otimes |u_2\rangle) = (|v_1\rangle, |v_2\rangle).(|u_1\rangle, |u_2\rangle),$$

which could be written in another notation as

$$\langle v_1 \otimes u_1 | v_2 \otimes u_2 \rangle = \langle v_1 | v_2 \rangle \langle u_1 | u_2 \rangle,$$

ENTANGLED QUANTUM STATES

The fundamental observation of Josza R., in "Entanglement and quantum computation", states that entanglement, not superposition, is the essential feature that empowers quantum computation, and is what gives other quantum technologies (such as quantum teleportation) their power. We will see in what follow what are the relations between superposition and entanglement.

For example, if we have two qubits with bases $\{|0\rangle_1, |1\rangle_1\}$ and $\{|0\rangle_2, |1\rangle_2\}$ respectively, the tensor product space has the basis

$$\{|0\rangle_1 \otimes |0\rangle_2, |0\rangle_1 \otimes |1\rangle_2, |1\rangle_1 \otimes |0\rangle_2, |1\rangle_1 \otimes |1\rangle_2\}$$

We can (conveniently) denote this basis as

$$\{|00\rangle, |01\rangle, |10\rangle, |11\rangle\}$$

More generally, if we have $n$ qubits to which we can apply common measurements, we will be working in the $2^n$-dimensional Hilbert space with basis

$$\{|00...00\rangle, |00...01\rangle, ..., |11...10\rangle, |11...11\rangle\}$$

A typical quantum state for an $n$-qubit system is $\sum_{i=0}^{2^n-1} a_i |i\rangle$ where $a_i \in \mathbb{C}, \sum |a_i|^2 = 1$, and $\{|i\rangle\}$ is the basis, with $i$ written as an $n$-bit binary number.

A classical (macroscopic) physical object broken into pieces can be described and measured as separate components. An $n$-particle quantum system cannot always be described in terms of the states of its component pieces. For instance, the state $|00\rangle + |11\rangle$ cannot be decomposed into separate states of each of the two qubits in the form $(a_1|0\rangle, b_1|1\rangle) \otimes (a_2|0\rangle + b_2|1\rangle)$.

This is because $(a_1|0\rangle, b_1|1\rangle) \otimes (a_2|0\rangle + b_2|1\rangle) = a_1 a_2 |00\rangle + a_1 b_2 |01\rangle + b_1 a_2 |10\rangle, b_1 b_2 |11\rangle$ and $a_1 b_2 = 0$ implies that either $a_1 a_2 = 0$ or $b_1 b_2 = 0$. States which cannot be decomposed in this way are called entangled states. These are states that don't have classical counterparts, and for which our intuition is likely to fail.

Particles are entangled if a measurement of one affects a measurement of the other. For example, the state $\frac{1}{\sqrt{2}}(|00\rangle + |11\rangle)$ is entangled since the probability of measuring the first bit as $|0\rangle$ is $1/2$ if the second bit has not been

measured. However, if the second bit has been measured, the probability that the first bit is measured as $|0\rangle$ is either 1 or 0, depending on whether the second bit was measured as $|0\rangle$ or $|1\rangle$, respectively. On the other hand, the state $\frac{1}{\sqrt{2}}(|00\rangle+|01\rangle)$ is not entangled. Since $\frac{1}{\sqrt{2}}(|00\rangle+|01\rangle) = |0\rangle \otimes \frac{1}{\sqrt{2}}(|0\rangle+|1\rangle)$, any measurement of the first bit will yield $|0\rangle$ regardless of measurements of the second bit. Similarly, the second bit has a fifty-fifty chance of being measured as $|0\rangle$ regardless of measurements of the first bit. Therefore, entanglement is a non-classical correlation between two quantum systems. It is most strongly exhibited by the maximally entangled states such as the Bell states for two qubits, and is considered to be absent in mixtures of product states ("separable" states). Often states that are not separable are considered to be entangled. However, nearly separable states do not exhibit all the features of maximally entangled states. As a result, studies of different types of entanglement are an important component of quantum information theory.

QUANTUM COMPUTING

This exponential growth in number of states, together with the ability to subject the entire space to transformations (either unitary dynamical evolution of the system, or a measurement projection into an eigenvector subspace), provides the foundation for quantum computing.

An interesting (apparent) dilemma is the energetic costs/ irreversibility of classical computing. Since unitary transformations are invertible, quantum computations (except measurements) will all be reversible by restricting them to unitary quantum transformations. This means that every quantum gate (on one or many qubits) implements a reversible computation. That is, given the output of the gate, it must be possible to unambiguously determine what the input was. Fortunately, there is a classical theory of reversible computation that tells us that every classical algorithm can be made reversible with an acceptable overhead, so this restriction on quantum computation does not pose a serious problem. It is something that must be kept in mind when proposing a specification for a quantum gate, however.

SIMPLE QUANTUM GATES

The quantum version of the classical *not* gate is represented by $\sigma_x$. It has the effect of negating the values of the computational basis. That is, using *ket* notation,

$$not(\alpha|0\rangle + \beta|1\rangle) = \alpha|1\rangle + \beta|0\rangle = \beta|0\rangle + \alpha|1\rangle$$

In vector notation this equation becomes: $\quad not\begin{pmatrix}\alpha\\\beta\end{pmatrix} = \begin{pmatrix}\beta\\\alpha\end{pmatrix}.$

Another effect of expressing the effect of *not* is by multiplying the vector by a matrix representing *not*:

$$not\begin{pmatrix}\alpha\\\beta\end{pmatrix} = \begin{pmatrix}0 & 1\\1 & 0\end{pmatrix}\begin{pmatrix}\alpha\\\beta\end{pmatrix} = \begin{pmatrix}\beta\\\alpha\end{pmatrix}$$

so we can identify the action of not with the matrix $\sigma_x = \begin{pmatrix}0 & 1\\1 & 0\end{pmatrix}.$

Similarly, we can find some useful single-qubit quantum state transformations. Because of linearity, the transformations are fully specified by their effect on the basis vectors. The associated matrix is also shown. They are known as the four the four Pauli gates.

$$I : \begin{matrix} |0\rangle \to |0\rangle \\ |1\rangle \to |1\rangle \end{matrix} \quad \begin{pmatrix} 1 & 0 \\ 0 & 1 \end{pmatrix}$$

$$\sigma_y : \begin{matrix} |0\rangle \to |1\rangle \\ |1\rangle \to -|0\rangle \end{matrix} \quad \begin{pmatrix} 0 & -1 \\ 1 & 0 \end{pmatrix}$$

$$\sigma_z : \begin{matrix} |0\rangle \to |0\rangle \\ |1\rangle \to -|1\rangle \end{matrix} \quad \begin{pmatrix} 1 & 0 \\ 0 & -1 \end{pmatrix}$$

Note that $I$ is the identity transformation (often called *nop* or no-operation), $\sigma_x$ is negation, $\sigma_z$ is a phase shift operation, and $\sigma_y = \sigma_z \sigma_x$ is a combination of both. One reason why the Pauli gates are important for quantum computing is that they span the vector space formed by all 1-qubit operators.

All these gates are indeed unitary. For example:

$$\sigma_y \sigma_y^\dagger = \begin{pmatrix} 0 & -1 \\ 1 & 0 \end{pmatrix} \begin{pmatrix} 0 & 1 \\ -1 & 0 \end{pmatrix}$$

Another very important gate is the Hadamard gate defined by the following transformation:

$$H : \begin{matrix} |0\rangle \to |0\rangle + |1\rangle \\ |1\rangle \to |0\rangle - |1\rangle \end{matrix}$$

Applied to *n* bits each in the $|0\rangle$ state, the transformation generates a superposition of all $2^n$ possible states.

$$H^{\otimes n}|0\rangle^{\otimes n} = \frac{1}{\sqrt{2^n}}((|0\rangle+|1\rangle)\otimes ... \text{n times} ... \otimes(|0\rangle+|1\rangle)) = \frac{1}{\sqrt{2^n}}\sum_{x=0}^{2^n-1}|x\rangle$$

Other then the Hadamard gate, we need to mention the T gate. It is sometimes referred to as the $\frac{\pi}{8}$ gate. It is represented by the following matrix:

$$T = \begin{pmatrix} 1 & 0 \\ 0 & e^{i\frac{\pi}{4}} \end{pmatrix} = e^{i\frac{\pi}{8}} \begin{pmatrix} e^{-i\frac{\pi}{8}} & 0 \\ 0 & e^{i\frac{\pi}{8}} \end{pmatrix}$$

Any arbitrary quantum gate could be synthesized from only a small minimal set. This universal set is the quantum analog of the universal gate (NOR or NAND) for classical systems. A universal set of operations is: H, X, T, and $C_{Not}$.

An important two-qubit operator is the *CNOT*. It is given as follows:

$$CNOT|00\rangle = |00\rangle$$
$$CNOT|01\rangle = |01\rangle$$
$$CNOT|10\rangle = |11\rangle$$
$$CNOT|11\rangle = |10\rangle$$

Classically, we can think of the *c-not* as flipping the second register if and only if the first register is set to 1. The transformation $C_{not}$ is unitary since $C_{not}^{\dagger} = C_{not}$ and $C_{not}C_{not} = I$. The $C_{not}$ gate cannot be decomposed into a tensor product of two single-bit transformations.

QUANTUM COMMUNICATION PROTOCOLS

In this section we examine two communication protocols which can be implemented using the tools we have developed in the previous sections. We will first present the superdense coding and next the quantum teleportation. No classical protocols can perform the same way. Those protocols try to carry out communication tasks between two parties. By convention, we will call them Alice and Bob.

Those protocols require two different communication channels: quantum and classical channel. Those channels carry quantum bits and classical bits respectively between two remote locations. We will use in the following the properties of entanglement to explore the advantage of quantum communication over its classical counterpart. Note that EPR pairs are maximally entangled qubit pair represented by the bell state $|\beta_{00}\rangle = \frac{1}{\sqrt{2}}(|00\rangle + |11\rangle)$.

The protocols require that initially Alice and Bob share an EPR pair (the sender has one half and the receiver has the other). Such a state would have been created ahead of time when the qubits were together. This entanglement becomes a resource which Alice and Bob use to achieve protocols such as the following.

*Super Dense Coding*

The idea is to send two classical bits of information by only sending one quantum bit. The protocol proceeds as follow:

If Alice wishes to send the bits 00 to Bob, she does nothing to her qubit (or equivalently, applies the identity gate *I*). If she wishes to send 01, she applies the $\sigma_x$ gate to her qubit. If she wishes to send 10, she applies the $\sigma_z$ gate, and if she wishes to send 11, she applies $i\sigma_y$ (if necessary review the Pauli gates from the previous section).

The resulting states of the 2-qubits are:

1. If classical bits are '00', apply *I*:

$$\frac{1}{\sqrt{2}}(|00\rangle+|11\rangle) \xrightarrow{I} \frac{1}{\sqrt{2}}(|00\rangle+|11\rangle) = |\beta_{00}\rangle$$

2. If classical bits are '01', apply $\sigma_x$:

$$\frac{1}{\sqrt{2}}(|00\rangle+|11\rangle) \xrightarrow{\sigma_x} \frac{1}{\sqrt{2}}(|01\rangle+|10\rangle) = |\beta_{01}\rangle$$

3. If classical bits are '10', apply $\sigma_z$:

$$\frac{1}{\sqrt{2}}(|00\rangle+|11\rangle) \xrightarrow{\sigma_z} \frac{1}{\sqrt{2}}(|00\rangle-|11\rangle) = |\beta_{10}\rangle$$

4. If classical bits are '11', apply $i\sigma_y$:

$$\frac{1}{\sqrt{2}}(|00\rangle+|11\rangle) \xrightarrow{i\sigma_y} \frac{1}{\sqrt{2}}(|01\rangle-|10\rangle) = |\beta_{11}\rangle$$

Until this stage, Alice just performed local unitary transformation on her qubit. Now Alice sends to Bob her qubit. From the above analysis, Bob is in possession of one of the above Bell states. Finally, he simply perform a measurement of the joint 2-qubit state in the Bell basis $\{|\beta_{00}\rangle,|\beta_{01}\rangle,|\beta_{10}\rangle,|\beta_{11}\rangle\}$. They key to super-dense coding is that they are orthonormal from each other and are hence distinguishable by a quantum measurement. That way, quantum mechanics allowed us to communicate two classical bits at the cost of one quantum bit.

*Quantum Teleportation*

Here is a protocol that allow Alice to communicate to Bob the state of a qubit $\psi = \alpha|0\rangle + \beta|1\rangle$. Classically, this would require communicating the value of α and β with an infinite amount of precision and therefore would turn out to be impossible. But, using one more time the properties of entanglement, we will illustrate how this would be possible.

Teleportation is a protocol which allows Alice to communicate the state of a qubit exactly to Bob, sending only two bits of classical information to him. Like superdense coding, teleportation requires that Alice and Bob initially share the bell state

$$|\beta_{00}\rangle = \frac{1}{\sqrt{2}}(|00\rangle + |11\rangle)$$

In brief, it could be stated as the following:

Teleportation works by pre-transmitting an EPR pair to the source and destination. The qubit containing the state to be "teleported" interacts with one half of this EPR pair creating a joint state space. It is then measured and only classical information is transmitted to the destination. This classical information is used to "fix up" the destination qubit

The equivalent quantum circuit is

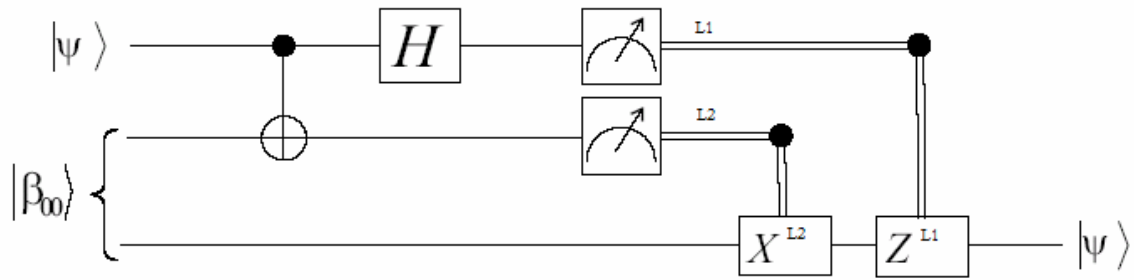

**Figure 4: Teleportation circuit**

Line 1 and 2 represents Alice's qubits, while line 3 (bottom line) represents Bob's qubit. The meters represent measurement devices, and the double lines coming out of them represent a classical channel (like a telephone) across which Alice can communicate the results of her measurements to Bob.

In this case, if she communicates the result of her measurement to Bob (over a classical channel), he can determine what operation he has to perform on his qubit to get the state $\psi$ out of it. Using our notational convention, mentioned in figure 1, Bob applies $Z^{L1}$ and $X^{L2}$ to his qubit consecutively. $Z^{L1}$ (or $X^{L2}$) means applying the Pauli gate $\sigma_z$ (or $\sigma_x$) conditioned on the Boolean value of the measurement of the first qubit (second qubit). After this transformation, Bob is guaranteed to have the state $\psi$ and so the state has been successfully teleported from Alice to Bob.

In the above presentation, the technical steps have been removed in order to emphasize on the protocol rather than the computations in the Hilbert space.

QUANTUM ALGORITHMS

Quantum algorithms are methods using quantum networks and processors to solve algorithmic problems. On a more technical level, a design of a quantum algorithm can be seen as a process of an efficient decomposition of a complex unitary transformation into products of elementary unitary operations (or gates), performing simple local changes.

The four main properties of quantum mechanics that are exploited in quantum computations are:

- Superposition
- Interference
- Entanglement
- Measurement

FACTORING

The publication of P. Shor's quantum algorithm for efficiently factoring numbers was the key event that stimulated many theoretical and experimental investigations of quantum computation. One of the reasons why this algorithm is so important is that the security of widely used public key cryptographic protocols relies on the conjectured difficulty of factoring large numbers.

The factoring problem requires writing a whole number N as a product of primes. Shor's algorithm solves this problem by reducing it to instances of order-finding problem efficiently.

To factor a number M, we choose a number $y < M$ with $\gcd(y,M) = 1$. We then find r, the order of y in the multiplicative group (mod M). If r is even, then $(y^{\frac{r}{2}}+1)(y^{\frac{r}{2}}-1) = (y^r - 1) \equiv 0 \pmod{M}$. Then $\gcd(y^r-1,M)$ is a non-trivial factor of M except when r is odd or $y^{\frac{r}{2}} \equiv -1 \pmod{M}$. This procedure produces a non-trivial factor of M with probability at least $1-\frac{1}{2^{k-1}}$, where k is the number of distinct odd prime factors of M. If we don't get a factor, we can choose a new y and repeat the process. By repeating the process, we can make our likelihood of success as close to one as we like.

Outline of Shor's algorithm for factoring a number *M*:

1. Choose an integer $y < M$ arbitrarily. If *y* is not relatively prime to *M*, we've found a factor of *M*. Otherwise apply the rest of the algorithm.

2. Let *n* be such that $M^2 \leq 2^n < 2M^2$. We begin with *n* qubits, each in state $|0\rangle$. We now apply the Hadamard transformation *H* to superpose all states:

$$H^{\otimes n}|0\rangle^{\otimes n} = \frac{1}{\sqrt{2^n}} \sum_{a=0}^{2^n-1} |a\rangle$$

3. Apply a transformation which implements raising to powers( mod M)

$$\frac{1}{\sqrt{2^n}} \sum_{a=0}^{2^n-1} |a\rangle \rightarrow \frac{1}{\sqrt{2^n}} \sum_{a=0}^{2^n-1} |a, f(a)\rangle$$

where $f(a) = y^a \pmod{M}$.

4. We want to find the period of the function $f(a) = y^a \pmod{M}$. We do that by measuring to find a state whose amplitude has the same period as $f$.

We measure the qubits of the state obtained from encoding $f(a)$. A random value $u$ is obtained. We don't actually use the value $u$; only the effect the measurement has on our set of superpositions is of interest. This measurement projects the state space onto the subspace compatible with the measured value, so the state after measurement is

$$C \sum_a g(a) |a, u\rangle,$$

for some scale factor $C$ where

$$g(a) \begin{cases} 1 & \text{if } f(a) = u \\ 0 & \text{otherwise} \end{cases}$$

Note that the $a$'s that actually appear in the sum, those with $g(a) \neq 0$, differ from each other by multiples of the period, and thus $g(a)$ is the function we are looking for. If we could just measure two successive $a$'s in the sum, we would have the period. Unfortunately the quantum world permits only one measurement.

5. Apply a quantum Fourier transform to invert the frequency. Shor's method uses a quantum version of the Fourier transform to find the period of the function $y^a \mod M$.

The quantum Fourier transform $U_{QFT}$ with base $2^n$ is defined by

$$|a\rangle \xrightarrow{QFT} \frac{1}{\sqrt{2}} \sum_{c=0}^{2^n-1} e^{2\pi i \frac{ca}{2^n}} |c\rangle$$

We apply the quantum Fourier transform to the state obtained by the measurement.

$$\sum_a g(a)|a\rangle \xrightarrow{QFT} \sum_c g(c)|c\rangle$$

Standard Fourier analysis tells us that when the period $r$ of $g(a)$ is a power of two, the result of the quantum Fourier transform is

$$C' \sum_j \rho_j \left| j \frac{2^n}{r} \right\rangle$$

where $|\rho_j| = 1$. When the period $r$ does not divide $2^n$, the transform approximates the exact case so most of the amplitude is attached to integers close to multiples of $\frac{2^n}{r}$.

6. Extract the period, which we expect to be the order of y (mod M).
7. Find a factor of $M$. When our estimate for the period, $q$, is even, we use the Euclidean algorithm to efficiently check whether either $y^{q/2}+1$ or $y^{q/2}-1$ has a non-trivial common factor with $M$.
8. Repeat the algorithm, if necessary.

In order for Shor's factoring algorithm to be a polynomial algorithm, the quantum Fourier transform must be efficiently computable. Shor developed a quantum Fourier transform construction with base $2^n$ using only $\frac{n(n+1)}{2}$ gates.

The construction makes use of two types of gates. One is the Hadamard transformation illustrated in previous sections. This is an exponential speed-up of the process since it can be accomplished in approximately $n^2$ operations rather than $n2^n$.

GROVER SEARCH ALGORITHM

Following is an outline of Grover's general search algorithm. If P(x) is a boolean function for $0 \leq x < N$, classical search algorithms take on the order of $\frac{N}{2}$ operations to find an item $x_0$ for which $P(x_0) = 1$. Grover's algorithm takes on the order of $\sqrt{N}$ operations. Grover's algorithm has been shown to be optimal for the general search problem. This is not an exponential speedup, but it is an improvement over the classical algorithms.

To solve the problem, Grover starts by setting a quantum register to a superposition of all possible items in the database. The quantum state contains the right answer, but if the register were observed at this point, the odds of picking the right answer would be as small as if one picked the item by random.

Grover's discovery involves a sequence of simple quantum operations on the register's state. He describes the process in terms of wave phenomena. "All the paths leading to the desired results interfere constructively, and the others ones interfere destructively and cancel each other out," Grover explains.

Grover's algorithm consists of the following steps:

1. Let n be such that $2^n \geq N$, and prepare a register containing a superposition of all $x_i \in [0...2^n - 1]$.

2. Apply a unitary transformation that computes $P(x_i)$ on this register:

$$U_P : \frac{1}{\sqrt{2^n}} \sum_{x=0}^{2^n-1} |x,0\rangle \to \frac{1}{\sqrt{2^n}} \sum_{x=0}^{2^n-1} |x, P(x)\rangle$$

For any $x_0$ such that $P(x_0)$ is true, $|x_0,1\rangle$ will be part of the resulting superposition, but since its amplitude is $\frac{1}{\sqrt{2^n}}$, the probability that a measurement produces $x_0$ is only $\frac{1}{2^n}$.

3. Change amplitude $a_j$ to $-a_j$ for all $x_j$ such that $P(x_j)=1$.

4. Apply inversion about the average to increase amplitude of $x_j$ with $P(x_j)=1$ and decrease other amplitudes.

5. Repeat steps 2 through 4 $\frac{\pi}{4}\sqrt{2^n}$ times.

6. Measure the last qubit of the quantum state, representing $P(x)$. Because of the amplitude change, there is a high probability that the result will be 1. If this is the case, the measurement has projected the state onto the subspace $\frac{1}{\sqrt{2^k}} \sum_{i=1}^{2^k} |x_i,1\rangle$ where k is the number of solutions. Further measurement of the remaining bits will provide one of these solutions.

An interesting feature of this algorithm is that repeating steps 2 through 4 $\frac{\pi}{4}\sqrt{2^n} \approx O(\sqrt{N})$ times is optimal. In particular, if the process is repeated more times, the probability of a successful measurement decreases back toward zero ...

# QUANTUM CRYPTOGRAPHY

*History of Quantum Cryptography*

Quantum cryptography was the first initiative to believe and to investigate in quantum network security. Stephen Weisner first proposed this idea in the early 1970's. However this idea wasn't eventually published untill 1983 in Sigact News, and following that, Bennet and Brassard published ideas of their owns in 1984 inspired from Weisner previous work. They produced the BB84[iv] which is considered the first quantum cryptographic (QC) communication protocol. This protocol has been experimentally implemented for over 50 kilometers using fiber optic cables and 1 kilometer in free space (Franson 1996). After nearly two decades as a laboratory curiosity, quantum key distribution (QKD) techniques are just now emerging as useful building blocks in highly secure networks.

*Current Research in Quantum Networking*

Recently on the 3rd of November 2003, MagiQ Technologies Inc., the quantum information processing company based in New York , launched the world's first commercially available quantum key distribution system. Called Navajo Secure Gateway, it supports secure key exchange at distances up to 120 km, a major technical accomplishment that makes very long secure spans possible via cascading devices[v]. This elaborated scheme does not use quantum effects to transmit secret data. Instead, it distributes secret keys based on quantum theory up to a rate of 1,000 keys a second. It have been proven that the risk to decrypt the data by an eavesdropper without the key is reduced to zero[vi].

On the other hand, a little bit North, a team from Boston, Harvard and BBN Technology sponsored by DARPA are undertaking a five year research program that consists of building and testing a highly secure quantum network, claiming to be a world premier[vii]. This ambitious project is supposed to perform extensive testing against sophisticated eavesdropping attacks.

Still in Boston, MIT, the leading university in computer science and engineering, is taking the initiative to develop designs for a quantum internet under the sponsor of the Army Research Office (MURI). Moreover, it[viii] is currently implementing those designs in parallel with corresponding protocols and applications. This design would allow the robust transmission of quantum information even in the presence of high levels of errors and loss.

At that time, across the Atlantic, three leading E-Security organizations in Geneva are joining forces to deploy what will be the first ever integrated Quantum Key Infrastructure. This initiative came after they joined hands with a trade organization in pledging to create the infrastructure necessary for worldwide distribution of unbreakable quantum keys at the ITU Telecom World 2003 conference in Geneva on the 15$^{th}$ of October 2003[ix].

BASIC IDEAS

After this encouraging overview, we would like to clarify that quantum cryptography currently does not present complete solutions for all cryptographic purposes. Instead it is used to complement secret-key based classical cryptosystems. More specifically, QC is known for its unconditional secure secret keys distribution such that it is seen as an excellent alternative to the Diffie-Hellman key exchange algorithm. Therefore we would like to rename *quantum cryptography* in this section as *quantum key distribution*, since this is all what is implemented so far.

*Heisenberg's uncertainty principle*

What arouses mostly the interest of researchers and investors in quantum key distribution is that its security is based on the laws of physics without making assumptions about the intractability of certain mathematical problems even when the eavesdroppers have access to unlimited computing power.

Heisenberg's uncertainty principle, quantum theory axiom, states that *we can not measure the system without perturbing it*. For cryptographic purposes, Heisenberg's principle could lead us to the following logic:

*No perturbation* ➔ *No measurement* ➔ *No eavesdropping* ➔ *No leak of information*

Stated in other words, quantum cryptography ensures that communications cannot be eavesdropped upon without introducing errors that can be readily detected by the receiver.

*Entanglement*

One additional feature in quantum mechanics that serves quantum cryptography is the *entanglement* between two quantum systems. It states that if two or more quantum systems that have interacted in the past may together share information in a form that they influence one another regardless of their spatial separation. A special case of entanglement principle is the EPR (Einstein, Poldolsky and Rosen) paradox; It says that as long as both photons remain unobserved, their properties remain indefinite, in a superposition of all states. But because of their common origin, the properties of the photons are entangled.

*Photons polarizations*

From the implementation point of view, the quantum bits (qubits) exchanged between Alice and Bob are encoded in the form of photons through a beam of light. At the best, each photon represents a single qubit. Suppose the photon's polarization chosen for encoding the bits of information is the following: vertical polarization ↕ for "1" and horizontal polarization ↔ for "0". Thus, the sequence of pulses ↕↔↕↕↔ corresponds to "10110". In measuring polarization of photons, we refer to a pair of orthogonal polarization as a basis.

In addition one could choose the diagonal polarization as a second basis such that / \ / / \ represents the string "10110". A pair of basis is said to be conjugate if the measurement of the polarization in the first basis randomizes the measurement in the second basis. For exemple, if you measure a horizontally or vertically polarized photon in the diagonal basis, you cannot determine any information about the initial polarization of the photon. Alternatively, the circular polarization could be used as a second basis.

In that context, the quantum communication channel for photons can be either free space or optical fiber – which could be the ones used in standard telecommunications. Thus, the communication channel is not really quantum whereas the information carriers are quantum.

# THE BB84 QKD PROTOCOL

*Two different protocols*

Currently, two different types of quantum key distribution were elaborated based on the two counter-intuitive features of quantum mechanics: uncertainty and entanglement (presented in the section above). The first type uses the polarization of photons to encode the bits of information and relies on the quantum randomness to keep Oscar from learning the secret key. The second type uses entangled photon states to encode the bits and relies on the fact that the information defining the key only "comes into being" after measurements performed by Alice and Bob (Artur Ekert -1991).

BB84 QKD protocol is actually the one which is theoretically and experimentally elaborated the most. This protocol belongs to the first type. We will investigate in the following the BB84 in details.

*Quantum transmission*

In the first step, Alice sends individual qubits to Bob in states chosen at random among the four states: ↔, ↕, / , \ which are identified as the polarization states "horizontal", "vertical", "45°", and "135°", respectively.

The individual qubits could be sent all at once or one after the other (much more practical), the only restriction being that Alice and Bob be able to establish a one-to-one correspondence between the transmitted and the received spins.

Next, Bob measures the incoming qubits in one of the two bases, chosen at random (using a random-number generator independent from that of Alice).

| Alice's random string | 1 | 0 | 1 | 0 | 0 | 1 | 1 | 0 | 1 | 1 | 0 | 1 | 0 |
|---|---|---|---|---|---|---|---|---|---|---|---|---|---|
| Alice's random basis: | + | X | + | + | X | X | + | + | X | X | + | + | X |
| Alice sends to Bob: | ↕ | / | ↕ | ↔ | / | \ | ↕ | ↔ | \ | \ | ↔ | ↕ | / |
| Bob measures with: | + | X | X | + | + | X | + | X | X | + | X | + | X |
| Bob's results: | ↕ | / | / | ↔ | ↕ | \ | ↕ | \ | \ | ↔ | \ | ↕ | / |

Consequently, from the probability theory, Bob will use a correct polarizer half of the time and an incorrect polarizer the other half of time. But 50% of the second half, he will still get the correct result and 50% he will get an incorrect result. Thus Bob will come up with the correct result 0.5 * 1 + 0.5 * 0.5 = 75% of the time. Under the perfect situation, uncertainty principle tells us neither Bob nor Oscar on average can obtain a measurement better than 75%. At this point, Bob and Alice will have perfectly correlated results whenever they use the same basis but uncorrelated results otherwise.

*Basis reconciliation*

Hence a straightforward error correction scheme is possible: Alice announces the bases to Bob and Bob announces the positions he measured in the right bases. If the state is compatible, they keep the bit; if not, they disregard it. The key shared at this point is called the *sifted* or *raw* key (note that it is not really *shared* since Alice's and Bob's versions are different due to noise and eavesdropping on the channel). In this way about 50% of the bit string is discarded.

| Valid data: | | | / | - | \ | | | \ | | | | / |
|---|---|---|---|---|---|---|---|---|---|---|---|---|

| Translated to key: | 1 | 0 | 0 | 1 | 1 | 1 | 1 | 0 |
|---|---|---|---|---|---|---|---|---|

At this stage of the protocol, Alice and Bob use a public channel for basis reconciliation. This is very common in crypto-protocols. This channel does not have to be confidential, only authentic.

Our analysis suggests that neither Alice nor Bob can decide which key results from the protocol. We can state that a secret key was established between them. Neither party in fact sent a secret key to the other. Indeed, it is the conjunction of both of their random number generator (or free will!) that produced the key.

POSSIBLE ATTACK MODEL…

*Tentative attack*

Essentially, to overcome the problem of eavesdropping, one might try to build protocols that, given Alice and Bob can only measure the Quantum Bit Error Rate (QBER), either provide them with a verifiably secure key or stop the protocol and inform the users that the key distribution has failed.

Consider a scenario were Oscar intercepts a qubit passing through the communication channel from Alice to Bob. The result of the interception would be detected by Bob since he will not receive the expected qubit, and announce this on a public channel so Alice can disregard it. Therefore Oscar's effect will only be to lower the bit rate, possibly to zero, but this does not give Bob any useful information. Therefore, for real eavesdropping Oscar must send a qubit to Bob. Ideally he would like to send this qubit in its original state, keeping a copy for himself.

## No cloning Theorem

The no-cloning theorem states that there is no unitary transformation that can take a state $|\psi\rangle_n |0\rangle_n$ into the state $|\psi\rangle_n |\psi\rangle_n$ for arbitrary $|\psi\rangle_n$.

The no-cloning theorem is an immediate consequence of linearity. If

$$U(|\psi\rangle|0\rangle) = |\psi\rangle|\psi\rangle \text{ and } U(|\phi\rangle|0\rangle) = |\phi\rangle|\phi\rangle \quad (1)$$

then it follows from linearity that

$$U(a|\psi\rangle + b|\phi\rangle)|0\rangle = aU|\psi\rangle|0\rangle + bU|\phi\rangle|0\rangle = a|\psi\rangle|\psi\rangle + b|\phi\rangle|\phi\rangle \quad (2)$$

But if U cloned arbitrary inputs, we would have

$$U(a|\psi\rangle + b|\phi\rangle)|0\rangle = (a|\psi\rangle + b|\phi\rangle)(a|\psi\rangle + b|\phi\rangle)$$

$$a^2|\psi\rangle|\psi\rangle + b^2|\phi\rangle|\phi\rangle + ab|\psi\rangle|\phi\rangle + ab|\phi\rangle|\psi\rangle, \quad (3)$$

which differs from (2) unless one of $a$ or $b$ is zero.

Therefore, a general quantum copy machine that copies any unknown state cannot be realized.

## Failed attack

As we have seen, perfect quantum copy machine cannot exist. Therefore Oscar cannot keep a perfect quantum copy. In classical information, copying is a fundamental process that is frequently used to an extent that the 'Fan-out' gate is usually omitted from, and is assumed to be a natural part of, a classical circuit. This is in sharp contrast with quantum information, where the fact that quantum states cannot be copied is a fundamental attribute. This major difference is one of the properties that make quantum information attractive, since the limitations introduced by the no-cloning theorem have a positive side:

it prevents Oscar from perfect eavesdropping and hence makes QC potentially secure.

## A MORE REALISTIC APPROACH

So far, all we have shown is that Alice and Bob can arrive at a shared key without publicly announcing any of the bits. But in real life there are always some errors due to noise in the channel and the equipments, and from Oscar who is trying to gain information.

We will briefly introduce some encountered technical problems. First, Alice and Bob's bits will defer due to real photon detectors. Second, actual photon emitters generate an average number, n, of photons per pulse of light. They can not reliably generate single photons. In addition, they can not maintain the same average number, n, each time which makes it difficult for Alice and Bob to agree on a one to one correspondence between the exchanged qubits.

On the other hand, Oscar will higher the quantum bit error rate (QBER) through intercepting the photons as they are transmitted from Alice to Bob. Since communication errors and eavesdropping cannot be distinguished, Alice and Bob have to assume that all discrepancies are due to Oscar in order to be on the safe side.

Therefore, Alice and Bob must apply some classical information processing protocols, like error correction and privacy amplification to their data. The first protocol is necessary to obtain identical keys and the second to obtain a secret key.

*Intercept-resend strategy*

This simple and even practical attack consists of Oscar intercepting, measuring and qubits to Bob. Since Oscar, like Bob, has no idea which basis Alice uses to transmit each photon, he too must choose bases at random for his measurements. He then resends to Bob another qubit in the state corresponding to his measurement. If luckily he chooses the correct basis, Alice and Bob will not notice his intervention and all will go well. But suppose he chooses the wrong basis. He will equally likely send back to Bob either polarization for the photon in his measurement.

Those equally likely previous scenarios necessarily happen since Oscar has no information about Alice's random-number generator (hence the importance of the generator being truly random)

*Analysis*

The correct rate in each case when Oscar is present:

|  | Bob Right Polarizer |
|---|---|
| Oscar Right Polarizer | 0.5 * 1 = 0.5 |
| Oscar Wrong Polarizer | 0.5 * 0.5 = .25 |

The correct rate in average is $0.5 + 0.25 = 0.75$, such that the error rate in average is $1 - 0.75 = 0.25$. In this case, Oscar gets 50% information whereas he leaves a 25% error rate in the sifted key. Alice and Bob can thus easily detect the presence of Oscar. We will assume in the following that Oscar perform two actions: opaque eavesdropping with probability of $\lambda$ and no eavesdropping with probability of $1-\lambda$. Thus Bob's row key will not agree with Alice's row key with probability of $0.25 * \lambda = 0.25 \lambda$. If, however, Oscar applies this strategy to only a fraction of the communication, say $\lambda = 10\%$, then the error rate will be only 2.5%, while Oscar's information will be 5%.

*Reconciliation (extension of the BB84)*

At this stage, this key contains errors. The errors are caused by technical imperfections, as well as possibly by Oscar's intervention. Reconciliation is the first classical information processing protocol performed on the sifted key to obtain identical keys between Alice and Bob. Since Oscar presumably listens to all public transmissions, Alice and Bob must reveal as little information as possible while still ensuring that they end up with identical keys. Error reconciliation like the base reconciliation is performed over a public channel. We will follow in the following the BBBSS[x] reconciliation procedure.

In order to minimize the information exposed to Oscar, Alice and Bob first agree on a random permutation of the bits in their sifted keys (to randomize the location of errors). Then, they split the resulting string into blocs of size *b*. The constant *b* is chosen experimentally (BBBSS implementation) rather theoretically such that it is unlikely to contain more than one error. Alice and Bob then compare the parity of each bloc. If they find a pair of blocks with mismatched parities, they continually bisect the block into smaller and smaller blocks, comparing parities each time, until the error is found. Finally, to ensure that Oscar learns nothing from this process, Alice and Bob discard the last bit of each block whose parity they disclose. At the end, if the resulting error-rate $e_{real} > e_{max}$ then Alice and Bob abort.

*Privacy Amplification (extension of the BB84)*

At this point, Alice and Bob possess identical strings, but those strings are not completely private. Oscar may have gained some information about them through intercept/resend. During the reconciliation phase, Oscar did not gain any information, since the last bit of each parity check block was discarded. However, some of his original information about specific bits may have been converted to information about parity bits.

Privacy amplification which was developed by Ueli Maurer and other classical cryptographers, while quantum cryptography was being developed, turned quantum cryptography into a practical technology for secure communications. It is a sort of cryptographic version of error correction, which allows Alice and Bob to start with similar shared random keys about which Oscar has some information and make shorter shared random keys which are identical and about which Oscar has (essentially) no information.

Alice announces to Bob the description of a randomly selected hashing function **f** from **N-K** bits to **N-K-L-R-S** secret bits. They each applies **f** on the reconciled key in order to get the final secret-key $\mathbf{X_f}$ in $\{0,1\}^{N-K-L-R-S}$.

THE B92

Since the security of QC relies on the inability of Oscar to distinguish unambiguously and without perturbation between the different states that Alice sends to Bob, Bennett noticed that only two non-orthogonal states are needed. The four different states used in the BB84 are more than really necessary for QC. Although two non-orthogonal states are enough, it is not very good in practice since one can unambiguously distinguish between them at the cost of some losses[xi]. In 1992, Bennett came up with B92[xii], quantum key distribution protocol, that is a slight modification of BB84 relying on our discussion above.

CONCLUSION

Quantum key distribution protocols achieve something that ordinary classical cryptography cannot. They allow Alice and Bob to generate and share random secret keys which exhibit very small error rate. They also allow Alice and Bob to estimate the level of eavesdropping and so try to reduce the error introduced by Oscar and to amplify the privacy of their shared random keys.

# BUILDING ON CURRENT INFRASTRUCTURE

Currently, quantum network security does not appear as an independent application that provides complete protocols for secure communication. However, quantum key distribution techniques go along with the well established internet technology. They are employed in conjunction with the public internet or, more likely, with private networks that employ the internet protocol suite, in order to build secure communication systems. We note that such private networks are currently in widespread use around the world with customers who desire secure and private communications, e.g. financial institutions, governmental organizations, militaries and so forth, and that a marriage of QKD technologies to these types of private network may prove both feasible and immediately appealing in certain contexts.

The following discussion briefly describes the system architecture proposed by a team from BBN Technologies, Boston University and Harvard University under the sponsorship of DARPA[xiii].

## SECURITY ARCHITECTURE

Today, secure communication between cryptographic gateways or more indeed between individual computers on the internet, is provided by the well defined architecture of IPsec. It specifies the protocols, algorithms, databases and policies required for secure communication. Therefore, it would be optimal if we marry QKD technology with the current well established internet security architecture. This joint effort would guaranty secure internet traffic via quantum cryptography.

The following figure resolves this basic setup into considerably more detail.

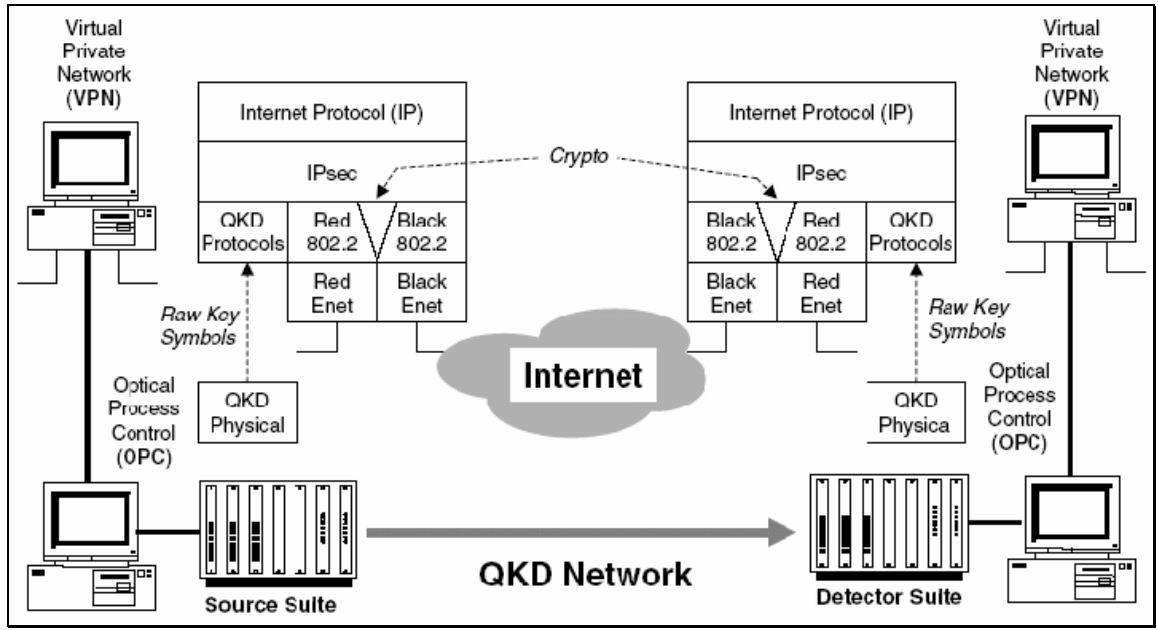

Figure 5: System architecture for a point-to-point QKD link

The basic concepts, however, are not difficult:

1. Two QKD endpoints establish communications via a dedicated fiber or wavelength for the quantum path, and via the internet for messaging;

2. The transmit side prepares and transmits raw keys, from which both sides come to agreement on a shared, secret key;

3. This secret key is then employed in the cryptographic gateway for protecting message traffic that will transit the internet within secured IPsec tunnels

Figure 2 provides a multi-layer approach for the QKD protocol explained in detail in the previous section. Those layers outline the degree of freedom each layer exhibit when looking for design alternatives.

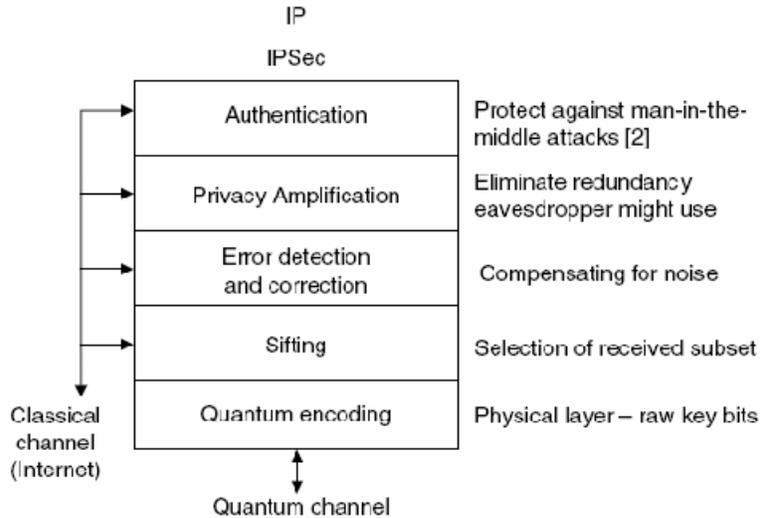

**Figure 6: Internal structure and functionality of QKD protocol suite.**

SYSTEM ARCHITECTURE

The following discussion briefly describes the system architecture of a quantum network as it evolves through three major stages, from a single, stand-alone QKD link supporting highly secure internet communications, through both trusted and untrusted QKD networks.

*System architecture of a point-to-point QKD link*

Figure 3 presents a simplified, block diagram of a point-to-point QKD link as it would most likely be deployed for secure networking, e.g. one that securely links a branch office to a corporate headquarters. Each enclave is typically a collection of one or more local Ethernets that connect to the public internet via specialized devices such as VPN gateway. That way, one needs to administer only a single device in order to establish or monitor external security for a given private enclave. These gateways are responsible for setting up security associations (and thus encrypted tunnels) with authorized distant gateway(s), for encrypting all local traffic before it is injected into the public network and for decrypting and authenticating traffic received from the public network before sending it onwards, in the clear, within the destination enclave. Given the nature of QKD, one would need two distinct communications paths: one for the cryptographic keys themselves, the other for the encrypted message traffic.

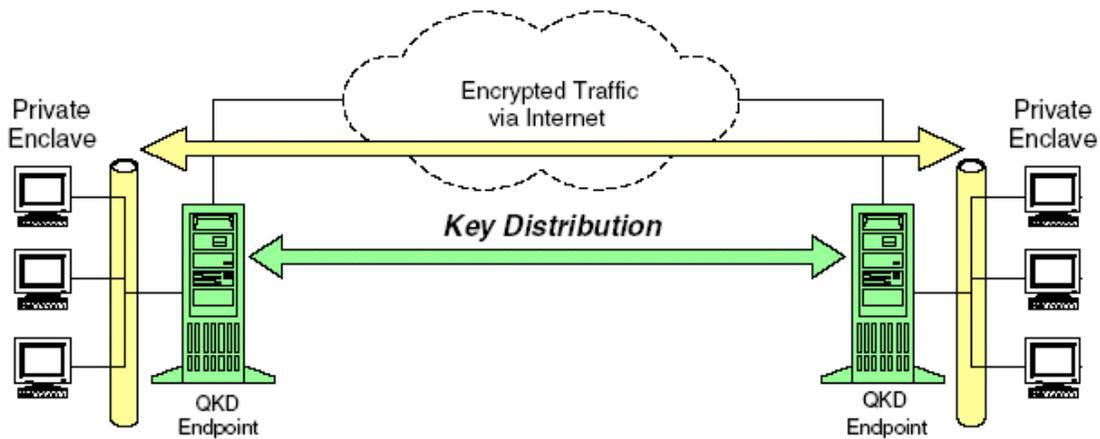

**Figure 7: Simplified block diagram of a point-to-point QKD link in context.**

*Drawbacks of the Point-to-Point QKD Architecture*

The above architecture suffers from striking drawbacks mainly due to the current technology. As discussed previously, fiber attenuation and error infiltrations limit terrestrial links to 50 km or less. Unfortunately, the point-to-point architecture is geographically constrained by the distance over which a single link may be operated. Moreover, isolated point-to-point links are subject to simple denial-of-service attacks such as active eavesdropping or cutting the fiber. Finally, in practice it may be prohibitively expensive to establish pair wise, dedicated point-to-point links between all private enclaves that wish to communicate with each other.

*System architecture of a trusted network*

Fortunately, these drawbacks can be attenuated by linking the QKD endpoints via a mesh of QKD relays or routers leading to a QKD network.

Such QKD networks can be built in several ways. In one variant, the QKD relays may transport only keying material but never message traffic. Thus after the various relays have established pair wise agreed-to keys along an end-to-end point, e.g. between the two QKD endpoints, they may employ these key pairs to securely transport a key 'hop by op' from one endpoint to the other, being one-time-pad encrypted and decrypted with each pair wise key as it proceeds from one relay to the next. In this approach, the end-to-end key will appear in the clear within the relays' memories proper, but will always be encrypted when passing across a link. Such a design may be termed a '*key transport network*'.

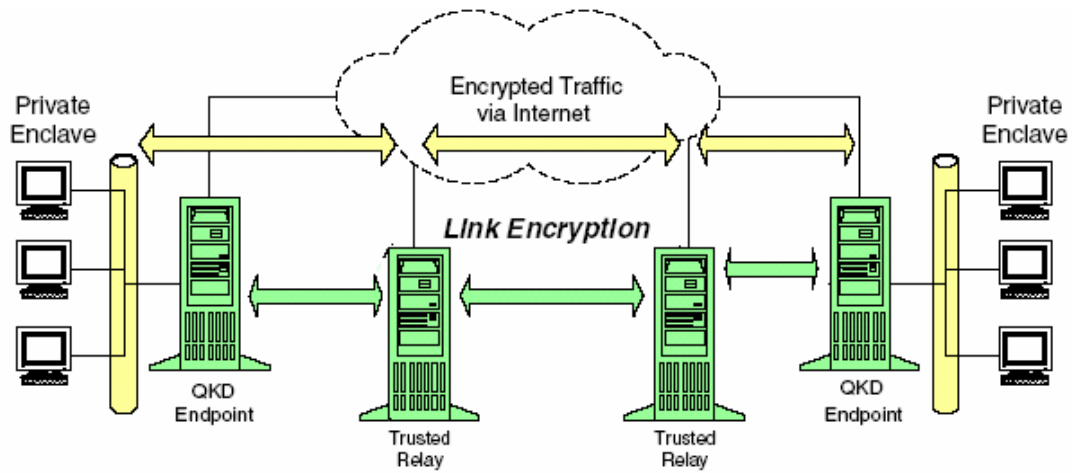

**Figure 8: QKD network with trusted relays and link encryption.**

In another variant, the QKD relays may transport both keying material and message traffic. Figure 4 illustrates this second variant, in which the relays are acting as internet-like routers with pair wise QKD mechanisms providing link encryption between the routers. In essence, each IP datagram of message traffic is encrypted once as it transits from the QKD endpoint to its first relay. Then it is decrypted, held in the clear in the relay's memory, and then re-encrypted with a second set of keys and sent onwards to the next relay. This operation proceeds, hop by hop, until the datagram is finally received at the destination endpoint and sent onwards to the attached private enclave. We note that this network differs from the standard definition of the internet by interposing a set of encrypted tunnels ('virtual links') between cooperating routers.

*Drawbacks of a Trusted Network Architecture*

The prime weakness of the above architecture is that the relays must be *trusted*. That is, since keying material and—directly or indirectly—message traffic are available in the clear in the relays' memories, these relays must be prevented from falling into an adversary's hands. Therefore, all users in the system must trust the network (and the network's operators) which provide all keys to their message traffic. Thus one should be careful if he deals with unusual sensitive traffic. However, fibers or free space links between the relays do not need such protection.

*System architecture of an untrusted network*

As in classical cryptography, an end-to-end approach is likely to provide the most satisfactory architecture for disentangling the users' keying material for secured traffic flows from the network that transports such flows. We present in the following an approach that introduces unamplified photonic switches into the QKD network architecture in order to provide end-to-end key distribution via a novel mesh of untrusted switches.

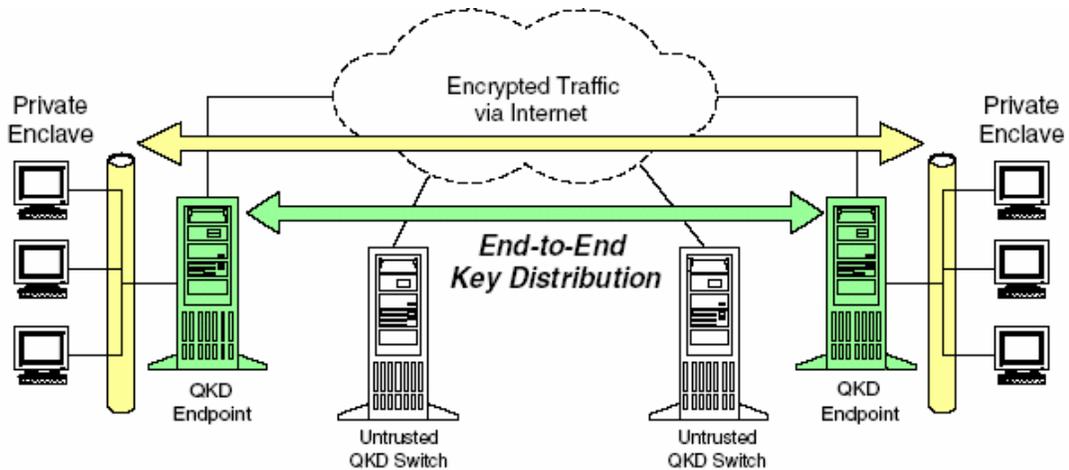

**Figure 9: QKD network with untrusted photonic switches.**

Untrusted QKD networks have different strengths and weaknesses than trusted QKD networks. Their main strength is that they provide truly end-to-end key distribution; QKD endpoints need not share any secrets with the key distribution network or its operators. This feature may be extremely important for highly secure networks. Their weaknesses appear significant, however. Unlike trusted relays, the untrusted switches cannot extend the geographic reach of a QKD network. In fact, they may significantly reduce the network's reach since each switch adds at least several dB losses to the photonic path. In addition, it will likely prove difficult in practice to employ a variety of transmission media within an untrusted network, since a single frequency may not work well along a composite path that includes both fiber and free space links. Untrusted networks may also introduce new vulnerabilities to traffic analysis.

On a cheerier note, the principal weakness in untrusted QKD networks—limited geographic reach—may potentially be countered by quantum repeaters. There is currently a great deal of active research aiming towards such repeaters, and if practical devices are ever achieved, they should slide neatly into the overall architecture of untrusted QKD networks to enable seamless QKD operations over much greater distances than are currently feasible.

QKD techniques can be married to standard internet technology in order to provide highly secure communications for practical use. Those designs elaborated above showed how to integrate both weak coherent and entangled QKD links with internet technology, and demonstrated QKD networks of both trusted (opto-electronic) and untrusted (passive photonic) switches. Such networks should be able to route around eavesdropping, noise and denial-of-service attacks on the QKD links.

# QUANTUM NETWORK SECURITY

In classical information one usually restricts oneself to binary representation of data, since using larger bases does not offer fundamentally additional tasks. Can we assume the same for quantum information? That is, can we make full use of the physical resources offered by quantum by limiting our information carries to qubits? It seems not.

## BYZANTINE AGREEMENT PROBLEM

Matthias Fitzi, Nicolas Gisin, and Ueli Maurer from Switzerland proposed a quantum solution for a slightly modified version of the Byzantine Agreement Problem using qutrits (i.e. a superposition of $|0\rangle$, $|1\rangle$, and $|2\rangle$), whereas no classical or qubit-based solution was found. The problem goes as follows:

The Byzantine army is divided into n divisions each commanded by a general, $m < n/2$ of which might be traitors. The divisions are camped around a city and the generals can only communicate in a one-to-one fashion using messengers. One of the generals, the commanding general, makes a detailed plan of an attack and wants to inform the others of it. Naturally the traitor generals are trying to prevent the loyal generals from reaching agreement on the plan of action. Note that the commanding general himself might be one of the traitors.

We will consider a modified version of the classical problem where the task is to find a protocol that achieves detectable broadcast, defined by that at end of protocol the generals agree on a commander's plan if everybody is loyal, otherwise either all loyal players agree on one plan of action or they abort the protocol.

The problem relates to coordinating several computers in a network where some might fail. However, detectable broadcast cannot be attained using classical channels. It is unproven whether there is a protocol to solve this problem using qubits, however it seems that there is none. For simplicity lets take n=3, and denote by S the commanding general, R0 & R1 the other two generals. We assume that the 3 players share many qutrits triplet's j, each in the Aharonov state that entangled is such a way that whenever the three qutrits are all measured, in the same basis, all three results differ. The following protocol was presented in Fitzi, Gisin, and Maurer's paper:

1. First, the sender S sends the bit x to be broadcast to the two receivers R0 and R1, using the classical channels. Let us denote x0 and x1 the bits received by R0 and R1, respectively. Next, the Sender S measures all his qutrits in the z-basis. Whenever he gets the result x, S sends the index j to both receivers. Accordingly, the players R0 and R1 receive each a set of indices, J0 and J1, respectively

2. Both receivers test the consistency of their data. For this they measure their qutrits in the z-basis. If all results with indices in Jp differ from xp, then player Rp has consistent data and he sets a flag $y_p = x_p$. If a set of data is inconsistent, then the player sets his flag to $y_p = ?$ (interpreted as inconsistent).

3. The two receivers send their flags to each other. If both flags agree then the protocol terminates with all honest players agreeing on x.

4. If $y_p = ?$, player Rp knows that the sender is dishonest. He concludes that the other receiver is honest and he simply accepts the bit he receives from him (If $y_0 = y_1 = ?$, then they both end with the "value" ?).

5. It remains only the interesting case that both receivers claim that they received consistent, but different, data. The strategy we propose then is that player R1 will not change his bit y1, unless player R0 convinces him that he did indeed receive the bit y0 from the sender in a consistent way. To convince his partner of his honesty, player R0 sends him all the indices k belongs to J0 for which he has the result 1-y0.

6. Receiver R1 now checks that he gets "enough" indices k from R0 such that

(a) "Almost all" indices k from R0 are not in R1's index set J1, and such that

(b) These k indices correspond to qutrits for which R1's results are "almost all" equal 2. If R0 indeed got an index set that is consistent with bit y0 then S holds y0, R0 holds 1-y0, and hence, R0's result must be a 2. If the test succeeds, player R1 changes his bit to y0, otherwise he keeps y1.

To see how the protocol works, consider the 6 possible cases {0 1 2, 0 2 1, 1 2 0, 1 0 2, 2 0 1, 2 1 0} that can occur when the three players measure their qutrits. If R0 receives the bit 0 then it is either that the qutrits are in the 0 1 2 or 0 2 1 and then R0 can prove that by announcing all the cases he obtained a value 1, for such cases R1 will have a value 2. If R0 pretends to have a 1 then he will be required to prove it by giving the indices that correspond to 2, however R0 cannot differentiate between cases 1 0 2 and 2 0 1. Thus approximately only half of the indices that R0 sends to R1 give a value of 1, then R1 realizes that R0 is cheating.

FINGERPRINTING

Let's move to a different problem in network security: Fingerprinting, a mechanism that arises in the study of communication complexity, is the problem of determining if two strings are equal as little communication and storage of information as possible.

The model of communication complexity considered is called the *simultaneous message passing* model, which was introduced by Yao[xiv]. In this particular model, Alice and Bob receive, respectively, and are not permitted to communicate with one another directly. Rather they each send a message to a third party, called the *referee*, who determines the output of the protocol based on the messages sent by Alice and Bob. The collective goal of the three parties is to cause the protocol to check if x = y, while minimizing the amount of communication required from Alice and Bob to the referee.

Newman and Szegedy[xv] proved that, classically, fingerprints of size $O(n^{1/2})$ bits are required to obtain a small probability of error if Alice and Bob do not have a prior shared secrete key. In quantum, Buhrman, Cleve, Watrous, and Wolf showed how to do this using $O(\log_2 n)$-qubits and proved that their method is nearly optimal.

The power behind quantum systems that allow this exponential save up is that quantum systems contain large sets of nearly orthogonal states. It is known that there are sets of $2^n$ states that are nearly orthogonal pair wise in $O(\log_2 n)$-qubit systems. Buhrman, Cleve, Watrous, and Wolf used these pairs to encode the strings:

$$|h_x\rangle = \frac{1}{\sqrt{m}} \sum_{i=1}^{m} |i\rangle |E_i(x)\rangle$$

Where $E_i(x)$ is the error correcting codes for our strings - these are just expanded version of n into m = cn, such that and such that the distance between distinct code words $E(x)$ and $E(y)$ is at least $(1 - \delta)$ m.

On the referee's side, comparison is done using the following quantum circuit:

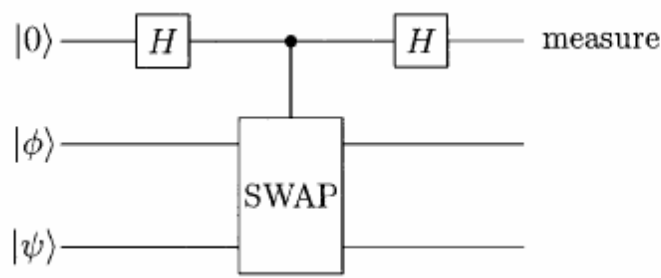

The circuit distinguishes between two states $|\Psi\rangle$ and $|\Phi\rangle$. If the two states are equal the circuit is guaranteed to output true, otherwise it outputs true with probability $(1+\delta^2)/2$. Thus we can repeat this protocol if the result keeps on being true until we are satisfied with the probability of success.

It is easy to see that $O(\log_2 n)$ is nearly optimal for quantum, given that any k-qubit quantum state can be specified with exponential precision using $O(k2^k)$ bits and since $O(n^{1/2})$ is the lower bound on size of the fingerprint for classical bits.

The possibilities and the limitations of quantum schemes are still unclear, the field is still immature and our intuition has yet to adapt to strange world of quantum mechanics. Perhaps the best demonstration of this is the history of quantum bit commitment:

Suppose that Alice has a bit b, a commitment scheme ensures to Bob that Alice cannot change the value of b, while it ensures to Alice that Bob knows the value of b only after a given time.

Several quantum schemes for bit commitment have been proposed in the literature and it was even claimed, by eminent researchers, that at least one is provably unbreakable. However, it was later shown that an EPR-type of attack - in which a dishonest Alice simply entangles the committed qubit and then later measure her half in an appropriate bases - render all proposed schemes insecure and was generalized to all non-relativistic quantum bit commitment scheme, a result known as Meyer, Lo and Chau theorem[xvi][xvii].

Kent[xviii] devised a quantum bit string commitment protocol, which is similar to bit commitment where the data being committed are strings of length n while only a subset of the string is recovered. The success of this protocol is counter-intuitive since classical reasoning incorrectly suggests that quantum bit string commitment implies quantum bit commitment.

Are there other undiscovered classically infeasible or inefficient tasks that quantum mechanics can help solve? We believe that there is plenty and we present one novel quantum mechanical scheme in the next section.

# QUANTUM CONTRACTS

The idea of Quantum Contracts (QCNTs) is motivated by Quantum Game Theory. In order to get insight into how quantum phenomena operate as a contract, we start by briefly discussing the Quantized Prisoner's Dilemma:

The payoff matrix of the classical Prisoners' Dilemma is shown bellow. The dilemma occurs since defection (D) is a strictly dominant strategy for both players. This leads both players to mutual defection (DD), which is a Nash equilibrium point, while this is significantly less rewarding than mutual cooperation (CC) which is Pareto optimal.

|            | Bob: C | Bob: D |
|------------|--------|--------|
| Alice : C  | (3, 3) | (0, 5) |
| Alice : D  | (5, 0) | (1, 1) |

The quantized version of the game[xix] involves an initial state $|\Psi_0\rangle = \hat{J}|CC\rangle$ then Alice and Bob encode their moves by applying unitary operators $U_A$ and $U_B$, as depicted bellow, the resulting state $|\Psi_f\rangle$ is measured and the corresponding payoff are taken from the payoff matrix.

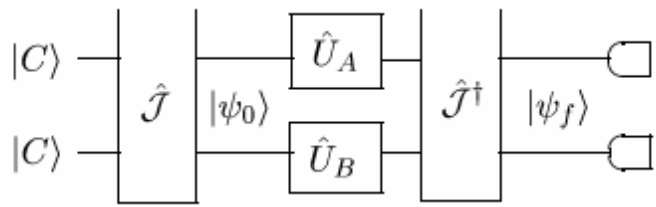

The role of *J* s is to entangle the states together enabling different strategies that correspond to relations between player's decisions. If we choose *J* to maximally entangle the qubits we discover a new Nash equilibrium strategy Q that effectively avoids the dilemma; when both players adopt Q, the resulting payoff is (3, 3).

In their study of Multiplayer Quantum Games, Benjamin and Hayden explained that the players escape the dilemma since "they can play cooperatively knowing that no player can successfully "defect" against the others. In this respect, quantum entanglement fulfills the role of a contract."[10].

We are now ready to introduce Quantum Contracts:

*Definition:* A Quantum Contract is a multiparty agreement that uses quantum mechanics to protect faithful participants by preventing an unfaithful participant from benefiting when violating.

Intuitively, QCNTs can be thought of as contracts that 'cancel' anytime it is attempted to violate them. QCNTs can be either one- or multi- sided. We first consider an example of a single sided contract:

Suppose Bob would like to use some information that Alice posses. Alice would like to satisfy Bob's demand, however she does not wish to allow him to use this information after a certain time. Let us further assume that Bob would benefit from keeping the information after that time.

If Alice trusts Bob then she can give him the information, however since Bob might keep the information, Alice rational choice depends on the following payoff matrix:

|           | Bob is F | Bob is NF |
|-----------|----------|-----------|
| Alice : C | (1, 3)   | (-5, 5)   |
| Alice : D | (0, 0)   | (0, 0)    |

If the probability that Bob is unfaithful is $p > 1/5$ then it is best for Alice not to send the information (D).

A one-sided QCNT scheme for this situation is for Alice to entangle the qubits representing the information to be sent to Bob with qubits she posses; then she will measure her qubits at the end of the allotted time. Effectively this will eliminate the -5 loss if Bob is unfaithful. The only possible scenarios left are shown bellow.

|           | Bob is F | Bob is NF |
|-----------|----------|-----------|
| Alice : C | (1, 3)   | (-5, 5)   |
| Alice : D | (0, 0)   | (0, 0)    |

The security of this QCNT is guaranteed by the non-cloning theorem and inability to perfectly distinguish between non-orthogonal states. Note that it is natural in the context of this problem that Alice allows limited amount of information to leak.

Next we give an example of a Double sided QCNT:

Consider a situation in which both Alice and Bob posses a collection qubits $\{b_1, b_2 \ldots b_n\}$ and $\{a_1, a_2 \ldots a_n\}$ respectively. Alice and Bob want each other's qubits. The problem with a classical two party deal is that if either Alice or Bob receive the information first there is no guarantee that they will complete their part of the deal.

For intuitive reasons we call this the hostage exchange problem. Classically this problem requires a mutually trusted third party to be solved.

We consider a slightly weaker variant of this problem, in which both Alice and Bob will need to use the qubits for at least time $t_1$ and $t_2$ respectively. The task is to find an exchange scheme that is valid till at least time $t = \min(t_1, t_2)$.

In a double sided QCNT scheme for this problem, Alice entangles the qubits in her possession obtaining $\{b_{1,1}, b_{2,1} \ldots b_{n,1}\}$ and $\{b_{1,2}, b_{2,2} \ldots b_{n,2}\}$, where each pair $b_{i,1}$ $b_{i,2}$ is a maximally entangled state having same coefficients as $b_i$. Bob does the same thing obtaining $\{a_{1,1}, a_{2,1} \ldots a_{n,1}\}$ and $\{a_{1,2}, a_{2,2} \ldots a_{n,2}\}$. Next, both Bob and Alice exchange the states $\{b_{1,1}, b_{2,1} \ldots b_{n,1}\}$ and $\{a_{1,1}, a_{2,1} \ldots a_{n,1}\}$.

Both Alice and Bob posses a method for testing for data integrity and they will then repeatedly use it to find if the data got corrupted by the opposing party. The contract can be clearly presented in a program:

While (t<t1) {

If (Test ($\hat{U}$ (qubits $a_1$)) = fail)) Then Measure (qubits $b_2$)

}

If (intention ≠ good) Measure (qubits $b_2$)

This previous code presents Alice's strategy; the conditional statement inside the loop continues to measures a known part of the evolved state in its bases. This will ensure Alice that Bob did not tamper with her qubits with high probability, note that this is sufficient since neither Bob nor Alice are willing to sacrifice their qubit for the sake destroying the others'.

Contracts are constructs that allow synergy among entities by mutual cooperation; it is beautiful to see that quantum mechanics enables such schemes to be implemented.

# INTERNET OF QUANTUM COMPUTERS

The quantum internet is analogous and complementary to the classical internet:

A hierarchical structure n-level network such that at level-0 the routers are connected to a number of quantum computers with groups of quantum routers at level-$i$ connected to level-$i+1$ quantum routers. A connection means the existence of a classical and a quantum channel.

On one level of abstraction, servers, which are quantum computers, can be thought of computing a function $Ux$ for some input $x$.

The non cloning theorem limits the ability of a server to distribute multiple copies of the result of his computation $Ux$.

Our first Protocol is designed to deal with this problem. The key idea here is that instead of sending $x$, Alice can alternatively send a program whose output is $x$ given a known input $y$.

The trade off in the protocol is between sending the program or sending a number of copies of $x$ that Alice can obtain by running her program several times.

It is sometimes beneficial for a computing device to compute his $Ux$ enough times to obtain the description of $Ux$ up to a desired degree of accuracy. Once $Ux$ is approximately known, one can use an algorithm by Long and Sun[xx] that efficiently initializes a quantum register with an arbitrary superposed state.

In addition Alice might prefer to send a program for its secrecy value, since the program then needs to be initialized with some input from both Alice and Bob, etc.

The next protocol is a means of entanglement distribution. As we have seen in the introduction, superdense coding and quantum teleportation imply a tradeoff in quantum communication in which a qubit and an EPR pair is equivalent to 2 classical bits and an EPR pair.

One way of transmitting messages between Alice and Bob is to simply teleport the qubits from Alice to her router at level-0 then iteratively up to a common level-$c$ then iteratively down to level-0 in Bob's neighborhood and finally to Bob.

An alternative way uses our entanglement distribution protocol. In which the mediating router prepares a bells state $\frac{1}{\sqrt{2}}(|00\rangle+|11\rangle)$ and Teleports the first qubit to Alice and the second to Bob. To our excitement the resulting qubits obtained by Alice and Bob are still in the Bell state.

In other words the protocol transfers the setting in which Alice and Bob share an entangled state with the router to a setting in which Alice shares an entangled state with Bob.

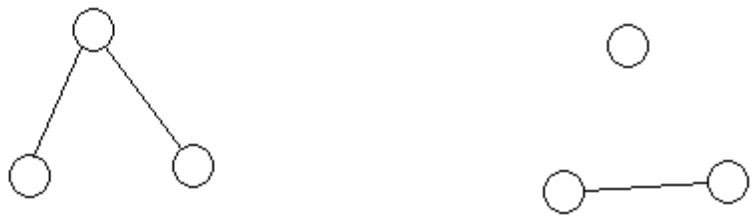

This is particularly powerful since it enables us to create virtual connections between parts of the quantum internet that do not have a direct quantum channels. The protocol will set up entangled states between Alice and Bob which can be then used to teleport the qubits directly.

We only give a hint on the third protocol since more work is being done on it and a related family of protocols. The setting is such that Alice can ensure that Bob performed a computation himself. This can be done when Alice entangles her data on which the computation to be performed with data that she posses, then Alice can check that Bob has not measured the qubit by measuring a certain subset on some bases.

We are currently working on crossing between Quantum contracts and the quantum internet; this is to be published soon.

What other discoveries awaits from our realization that information is physical? It is proven that if quantum mechanics turns out to be non-linear then it is straightforward that BQP=NP. What if we consider General Relativity? For one it tells us that computation is best done in the lowest density regions of space (?), what if time loops are allowed? What about string theory?

These are all intriguing questions but for now, our minds are allowed to relax… It will be a long time before our Quantum Computers seem too slow...